\documentclass[prd,nofootinbib,preprint,superscriptaddress,twocolumn,10pt]{revtex4}
\pdfoutput=1

\usepackage{amsmath,amssymb}
\usepackage{epsfig}
\usepackage{graphicx}
\usepackage[usenames,dvipsnames]{color}
\usepackage{subfigure}
\usepackage{slashed}
\usepackage[colorlinks,citecolor=blue]{hyperref}
\usepackage{color}

\begin{document}
	\title{Unified origin of dark matter self-interactions and low scale leptogenesis}

	\author{Debasish Borah}
	\affiliation{Department of Physics, Indian Institute of Technology Guwahati, Assam 781039, India}
	\author{Arnab Dasgupta}
	\affiliation{Pittsburgh Particle Physics, Astrophysics, and Cosmology Center, Department of Physics and Astronomy, University of Pittsburgh, Pittsburgh, PA 15206, USA}
	\author{Satyabrata Mahapatra}
	\affiliation{Department of Physics, Indian Institute of Technology Hyderabad, Kandi, Sangareddy 502285, Telangana, India}
	\author{Narendra Sahu}
	\affiliation{Department of Physics, Indian Institute of Technology Hyderabad, Kandi, Sangareddy 502285, Telangana, India}
	
	\begin{abstract}
		We propose a novel and minimal framework where a light scalar field can give rise to dark matter (DM) self-interactions, while enhancing the 
		CP asymmetry required for successful baryon asymmetry of the Universe via leptogenesis route. For demonstration purpose we 
		choose to work in a scotogenic seesaw scenario where the lightest among the right handed neutrinos (RHN), introduced for generating light neutrino masses 
		radiatively, play the role of DM while the heavier two can play non-trivial roles in generating DM relic as well as lepton asymmetry. While dark 
		matter self-interactions mediated by an additional singlet scalar can alleviate the small scale issues of cold dark matter paradigm, 
		the same scalar can give rise to new one-loop decay processes of heavy RHN into standard model leptons providing an enhanced contribution to CP asymmetry, 
		even with sub-TeV scale RHN mass. The thermally under-abundant relic of DM due to large annihilation rates into its light mediator receives a 
		late non-thermal contribution from a heavier RHN. With only five new particles involved in the scotogenic seesaw, each having non-trivial roles in 
		generating DM relic and baryon asymmetry, the model can explain non-zero neutrino mass while being verifiable at different experiments related to 
		DM direct detection, flavour physics and colliders. The mechanism we demonstrated here by using a scotogenic seesaw scenario is also 
		applicable to other models. 
	\end{abstract}	
	\maketitle
	
	\section{Introduction}
	\label{intro}
	Self-interacting dark matter (SIDM) is a promising alternative to the standard cold dark matter (CDM) paradigm in the light of the small scale issues 
	like too-big-to-fail, missing satellite and core-cusp problems faced by the latter \cite{Spergel:1999mh, Tulin:2017ara, Bullock:2017xww}. The required 
	self-interaction, often quantified in terms of cross section to dark matter (DM) mass as $\sigma/m \sim 1 \; {\rm cm}^2/{\rm g} \approx 2 \times 10^{-24} \; {\rm cm}^2/{\rm GeV}$ \cite{Buckley:2009in, Feng:2009hw, Feng:2009mn, Loeb:2010gj, Zavala:2012us, Vogelsberger:2012ku}, can be naturally realized in scenarios where 
	DM has a light mediator. In such a scenario, one can achieve velocity dependent DM self-interactions in order to solve the small scale issues while being 
	consistent with standard CDM properties at large scales \cite{Buckley:2009in, Feng:2009hw, Feng:2009mn, Loeb:2010gj, Bringmann:2016din, Kaplinghat:2015aga, Aarssen:2012fx, Tulin:2013teo}. The strong coupling of DM with its light mediators also leads to large DM annihilation rates, generating under-abundant 
	relic irrespective of thermal or non-thermal origin of DM. This usually requires a hybrid setup for DM production, beyond the most minimal ones.

	Similarly, the origin of baryon asymmetry in the universe (BAU) has been an unsolved mystery. Among several possible mechanisms of generating this asymmetry dynamically, baryogenesis via leptogenesis \cite{Fukugita:1986hr, Davidson:2008bu} has been a very popular one as it can connect to the origin of light neutrino mass via seesaw mechanism. However, in typical seesaw models, the scale of leptogenesis remains very high if such asymmetries are generated from decay \cite{Davidson:2002qv}, keeping it away from the reach of any direct experimental probe. However, it is worth mentioning that, lepton asymmetry can also be generated from oscillations \cite{Akhmedov:1998qx, Asaka:2005pn, Abada:2018oly, Drewes:2021nqr} where the scale of leptogenesis in minimal seesaw model can be as low as sub-GeV scale. In scenarios where lepton asymmetry is generated from decay, introduction of additional fields on top of the ones required to implement a generic seesaw model of neutrino mass, can alleviate such strong lower bound on the scale of leptogenesis \cite{LeDall:2014too, Alanne:2018brf}. Even in such leptogenesis from decay type scenarios, there is another way to have TeV scale leptogenesis by resonant enhancement of the CP asymmetry, known as the resonant leptogenesis \cite{Pilaftsis:1998pd} with fine-tuned mass splitting between decaying particles.
	
\begin{figure}[h!]
			\centering
			\includegraphics[scale=0.41]{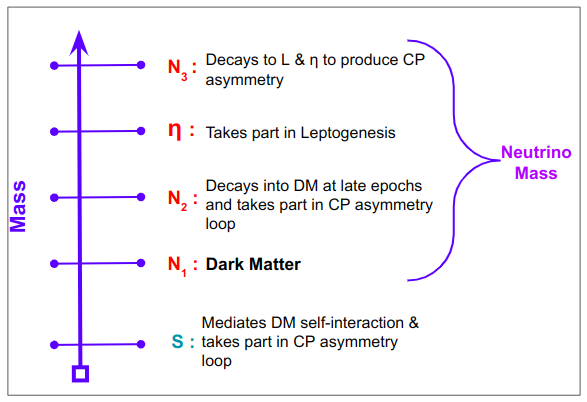}
			\caption{BSM particle spectrum and their roles in our setup.}
			\label{bsm}
		\end{figure}

	In this paper, we point out a novel connection between the SIDM and BAU, generated via a low scale leptogenesis route, thus unifying the origin of three 
	unsolved mysteries: SIDM, BAU and sub-eV neutrino mass origin at TeV scales. We propose a minimal framework where the fields taking part in one-loop decay of 
	a heavy right handed neutrino (RHN)into standard model (SM) leptons, leading to sufficient CP asymmetry required for leptogenesis, also address the issues of velocity dependent DM self-interactions as well as correct relic generation. We show that by promoting one of the heavy RHNs introduced in generic 
	seesaw models to play the role of DM by tuning or forbidding its couplings with leptons and introducing an additional light scalar field interacting with these 
	heavy RHNs including DM can lead to such a realization of unifying SIDM and low scale leptogenesis. In order to illustrate the connection quantitatively, we 
	choose to work in the framework of a radiative seesaw model where the desired interactions and DM stability can be ensured naturally due to the presence of 
	an unbroken discrete symmetry. To be more specific, we consider the addition of three RHN ($N_i, i=1,2,3$), a scalar doublet $(\eta)$ which are odd under an in-built $Z_2$ symmetry, similar to the minimal scotogenic model \cite{Ma:2006km}. In addition, we have a singlet scalar $(S)$ which is $Z_2$ even, similar to the SM particles. While $N_1$ with strong coupling to $S$ can give rise to self-interacting DM, its relic deficit can be filled by late decay of $N_2$. On the other hand, both $S$ and $N_2$ can take part in generating sufficient CP asymmetry due to decay of $N_3$ into leptons and $\eta$. In addition, all three RHNs and $\eta$ can give rise to radiative origin of light neutrino masses. The mass spectrum and roles of these new particles are summarized schematically in Fig. \ref{bsm}. We show that 
	this minimal framework is consistent with required self-interacting DM phenomenology as well as TeV scale leptogenesis while being verifiable at experiments 
	like charged lepton flavour violation (CLFV), DM direct detection. The mechanism, we demonstrated here by using a scotogenic seesaw, can also 
	be implemented in canonical seesaw scenarios but at the cost of ad-hoc fine-tunings of parameters to ensure DM stability. The details are under consideration 
	and will be reported elsewhere~\cite{Borah_in_prep}. It is worth mentioning that in an earlier work \cite{Mohanty:2019drv}, a scenario of scalar singlet SIDM and low scale leptogenesis was studied. However, the SIDM did not have any light mediator and its relic could be generated purely from thermal process by $3\rightarrow 2$ processes. However, such SIDM scenarios without light mediators do not have velocity dependent self-interactions required to solve the diversity problem mentioned earlier. Also the presence of light mediators in our scenario gives rise to a large $2 \rightarrow 2$ annihilation of DM requiring a hybrid setup for generating its relic. This makes our scenario very different from \cite{Mohanty:2019drv} in terms of relic generation, self-interaction of DM apart from other model details like origin of light neutrino masses etc. In another related work \cite{Kamada:2020buc}, maximally self-interacting DM scenario was connected to the origin of baryon asymmetry via a composite asymmetric DM setup. Once again, the generation mechanism of DM relic and lepton asymmetry in our work is very different from such scenarios, as we will discuss below.
	
	\section{The Framework}\label{sec2}
	The relevant Lagrangian for the minimal framework consisting of the particles mentioned above can be written as

		
		\begin{eqnarray}
			\label{lag}
			\mathcal{L} &=&\mathcal{L}_{\rm SM} +\overline{N_{i}}(i\gamma^\mu \partial_{\mu})N_{i}-\frac{1}{2}M_{N_{i}}\overline{N_{{i}}^c}N_{{i}} \nonumber \\
			&-& Y_{\alpha i} \overline{L_{\alpha}} \Tilde{\eta} N_{i} - y_{ij} S \overline{N_{{i}}^c}N_{{j}} + {\rm h.c.}
			+ \mathcal{L}_{\rm scalar}\,,
		\end{eqnarray}
		where $L,H$ are lepton and Higgs doublets of the SM while $\mathcal{L}_{\rm SM}, \mathcal{L}_{\rm scalar}$ are the SM Lagrangian, new scalar Lagrangian of the model respectively. The relevant terms of the scalar potential involving $\eta$ and $S$ are given  by
		\begin{eqnarray}
			V(H,\eta , S)& \supset & \frac{\lambda^{''}_{H \eta}}{2}\big[(H^\dagger \eta)^2+(\eta^\dagger H)^2\big]
			\\ \nonumber 
			& +& \mu_{S\eta} S \eta^{\dagger} \eta + \mu_{SH} S H^{\dagger} H
		\end{eqnarray}
		The singlet scalar can have quartic couplings with doublet scalars also, but since it does not acquire any vacuum expectation value (VEV) above 
		electroweak symmetry breaking (EWSB), only trilinear interactions are important for our analysis. 
		Since both the bare mass matrix of RHN and their Yukawa coupling matrix with singlet scalar are not necessarily diagonalizable simultaneously, we write the bare mass term in the diagonal basis whereas the coupling to the singlet scalar can be off-diagonal as well. Therefore, tuning the coupling of $N_{1}$ with $N_{2}, S$ small, we can have late decay of $N_{2}$ into DM. This brings the DM relic back to the correct ballpark which is otherwise under-abundant due to strong annihilation into light scalar mediator $S$. The mixing between the singlet scalar $S$ and the SM Higgs paves a way to detect DM at terrestrial laboratories such as CRESST-III and XENON1T. On the other hand, the off-diagonal coupling of $N_3$ with $N_2, S$ and trilinear coupling of $S$ with $\eta$ play crucial roles in generating a large CP asymmetry even for TeV scale mass of $N_3$.

		The light neutrino masses are generated at one-loop~\cite{Ma:2006km} and depends upon Dirac Yukawa couplings, masses of loop particles as well as the quartic coupling $\lambda''_{H \eta}$. Since the singlet scalar is assumed not to acquire VEV, we do not get additional one-loop diagrams for neutrino mass. Neutrino mass vanishes in the limit of $\lambda''_{H \eta} \to 0$ which corresponds to degenerate 
		neutral scalar and pseudoscalar (from $\eta$) masses. Thus, apart form the Yukawa couplings ($Y_{i \alpha}$) and RHN masses, the quartic coupling ($\lambda''_{H \eta}$) 
		also plays a significant role in neutrino mass generation. Constraints from neutrino data can be incorporated via the usual Casas-Ibarra parametrisation \cite{Toma:2013zsa}. Tuning the quartic coupling $\lambda''_{H \eta}$ to be small, one can enhance the Dirac Yukawa couplings. While this can have interesting implications for leptogenesis as well as DM phenomenology, such large Yukawa couplings can also lead to enhanced CLFV decays like $\mu \to e \gamma$ occurring at one-loop level which can be constrained from experimental data \cite{TheMEG:2016wtm}.

	\section{Dark Matter Phenomenology}\label{sec3}
	\begin{figure*}[t]
		\centering
		\begin{tabular}{cc}
			\includegraphics[scale=0.52]{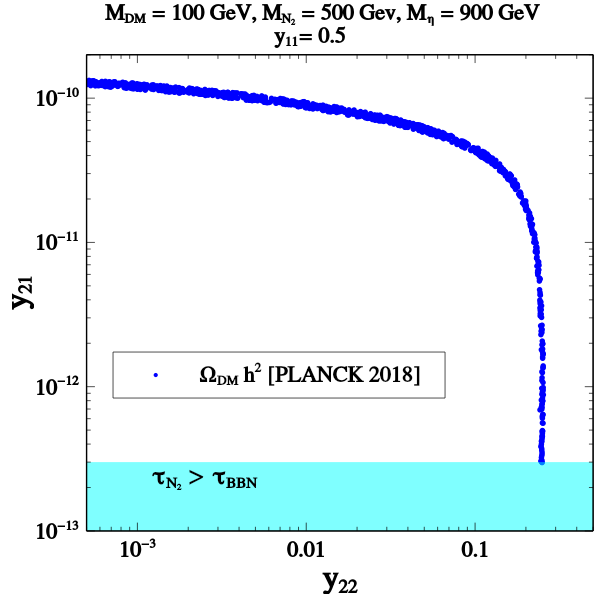}
			\includegraphics[scale=0.3]{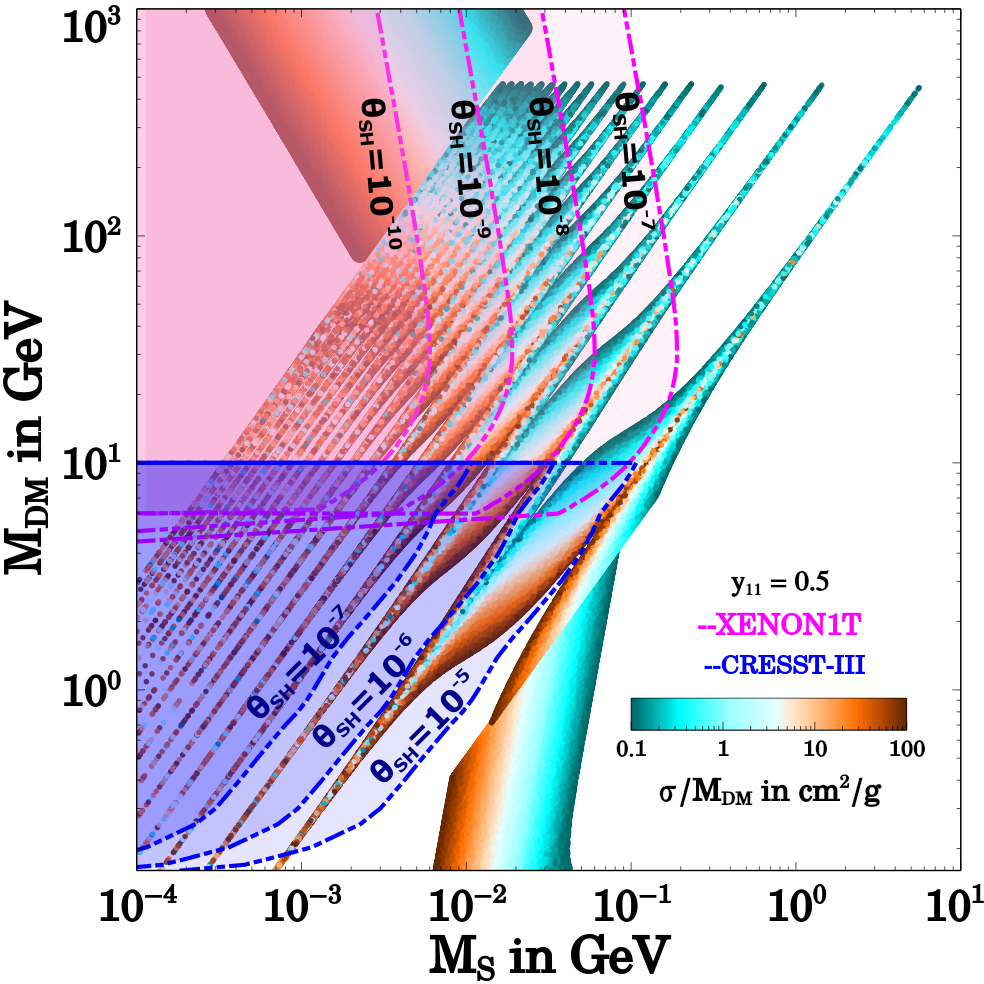}
		\end{tabular}
		\caption{{Left panel: Allowed parameter space in $y_{21}-y_{22}$ plane consistent with correct DM relic. Right panel: Constraints from DM direct detection in the plane of DM mass $(M_{\rm DM})$ versus mediator mass $(M_S)$ for self-interaction.}}
		\label{dmrelic}
	\end{figure*}
	The DM candidate is identified as the lightest RHN $N_1$ and its relic can be generated via a hybrid of thermal and non-thermal mechanisms \cite{Dutta:2021wbn, Borah:2021pet, Borah:2021rbx}. While all heavier $Z_2$ odd particles can decay into $N_1$, we consider only $N_2, \eta$ decay in our calculations for simplicity, so that $N_3$ decay can be dealt with separately in the context of leptogenesis only. Note that, we could have chosen the mass spectrum (Fig. \ref{bsm}) such that $\eta$ is the next to lightest $Z_2$-odd particle. However, due to large electroweak gauge interactions, its freeze-out abundance remains suppressed in the low mass regime and hence its late decay into DM may not be sufficient to get desired relic of the latter.

	We solve the relevant coupled Boltzmann equations numerically and find the evolution of relevant species for DM relic generation which is given in Appendix~\ref{dm_BE}. The DM is assumed to be produced in equilibrium in the early universe by virtue of its Yukawa couplings followed by its freeze-out resulting in an under-abundant relic primarily due to its large annihilation rates into light scalar $S$. The late decay of $N_2$ into DM can fill this deficit. Similar behaviour is obtained even if DM is assumed to be produced non-thermally from the SM bath at early epochs, followed by its annihilation into $S$, leaving a subdominant relic. Since $\eta$ can also decay into $N_2$ and leptons with much larger Yukawa coupling, its contribution to $N_1$ relic remains sub-dominant. The evolution of the comoving number densities DM and other relevant species ($N_2$ and $\eta$) has been shown in Fig.~\ref{dm_relic} for two benchmark points of the parameters. Please refer to Appendix.~\ref{dm_BE} for more details. On the left panel of Fig. \ref{dmrelic}, the parameter space in respective Yukawa couplings consistent with observed DM relic abundance is shown. While $y_{11}$ is large for self-interaction criteria, $y_{21}$ is tiny in order to have late decay of $N_2$ into DM. The Yukawa coupling $y_{22}$, on the other hand, decides the freeze-out relic of $N_2$ which later gets converted into DM relic\footnote{For chosen mass of $N_2$, its Yukawa coupling with SM leptons are more suppressed and hence play sub-dominant role in its freeze-out.}. When $y_{22}$ is small, then $N_2$ decouples from the  thermal bath early leaving a large freeze-out abundance. Therefore, in order to get correct DM relic, the decay of $N_2$ must occur at an early epoch so that subsequent DM annihilations into scalars can be effective enough to bring DM abundance within observed limits. For larger $y_{22}$, the freeze-out abundance of $N_2$ will be smaller and hence its decay into DM should occur at a relatively later epoch so that DM annihilations into scalars are no longer effective. This is reflected from the correlation between $y_{21}$ and $y_{22}$ shown on the left panel of Fig. \ref{dmrelic}.

	The DM has large elastic self-scattering due to its coupling with light mediator $S$. The corresponding non-relativistic DM scattering can be well described by the attractive Yukawa potential $V(r)= \frac{y^2_{11}}{4\pi r}e^{-M_{S}r}$. Calculating the quantum mechanical self-interaction cross sections and constraining $\sigma/M_{\rm DM}$ in the required range from astrophysical observations at different scales, such as dwarfs, low surface brightness (LSB) galaxies and clusters~\cite{Kaplinghat:2015aga,Kamada:2020buc}, we get the allowed parameter space of the model in the plane of DM mass $M_{\rm DM}$ and mediator mass $M_S$, shown in the right panel plot of Fig. \ref{dmrelic}. Thus, we see that the model can explain the astrophysical observation of velocity dependent DM self-interaction appreciably well. Here it is worth mentioning that calculations need to be performed much beyond the perturbative limit in order to obtain the whole parameter space shown and three different regimes can be distinguished depending on mass of the mediator and DM as well as relative velocity and interaction strength. These are the Born regime ($y^2_{11} M_{\rm DM}/(4\pi M_S) \ll 1,  M_{\rm DM} v/M_{S} \geq 1$) which is the lower region of the parameter space, the classical regime ($y^2_{11} M_{\rm DM}/4\pi M_S\geq 1$) corresponding to the upper end of the parameter space and and the region sandwiched between these two regions with sharp spikes is the resonant regime ($y^2_{11} M_{\rm DM}/(4\pi M_S) \geq 1, M_{\rm DM} v/M_{S} \leq 1$). These spikes appear because of quantum mechanical resonances and anti-resonances in the self interaction cross-section due to (quasi-)bound states formation in the attractive potential~\cite{Tulin:2013teo}. For details, please see Appendix~\ref{self_int}. In the same plot, we also show the contours for direct detection constraints. Direct search experiments like CRESST-III~\cite{Abdelhameed:2019hmk} and XENON1T \cite{Aprile:2018dbl} put severe constraints on the model parameters. The blue and magenta colored contours denote exclusion limits from CRESST-III~\cite{Abdelhameed:2019hmk} and XENON1T \cite{Aprile:2018dbl} respectively for different mixing angles $\theta_{SH} > 10^{-11}$. The region to the left of each contour is excluded for that particular mixing angle. It can be seen that $\theta_{SH}<10^{-11}$ is disfavoured for all $M_S \sim 10$ MeV. Note that $\theta_{SH}$ also has upper bound by invisible Higgs decay (as the singlet scalar is typically lighter than the Higgs mass), while a lower bound on $\theta_{SH}$ can be obtained by considering $S$ to decay before the big bang nucleosynthesis (BBN) epoch, {\it i.e.} $\tau_S <  \tau_{\rm BBN}$.

	\section{Leptogenesis}
	\label{sec3b}
	While the minimal scotogenic model by itself can give rise to successful leptogenesis at low scale \cite{Hambye:2009pw, Racker:2013lua, Hugle:2018qbw, Borah:2018rca, Mahanta:2019sfo}, the chosen hierarchy of $Z_2$-odd particles in our setup will lead to high scale leptogenesis \cite{Mahanta:2019gfe, JyotiDas:2021shi}. Therefore, we make use of the singlet scalar coupling to explore the possibility of low scale leptogenesis. Since singlet scalar couplings do not go into the neutrino mass, we have more freedom in generating a large CP asymmetry. In addition, the trilinear term of singlet scalar with $\eta$ (which assist in enhancing CP asymmetry) does not face stringent constraints like the one with SM Higgs. Since $N_{1,2}$ are lighter than $\eta$ they do not contribute directly to leptogenesis except being part of loop diagrams. While $N_3$ decay loop can contain $L, \eta, N_{1,2}$ alone without $S$, the corresponding CP asymmetry will be suppressed for TeV scale $N_3$ due to suppression in Yukawa couplings from the requirement of neutrino mass as well as sub-dominant washouts.

	The leptonic CP asymmetry $(\epsilon)$ expressions for the minimal scotogenic model can be found in \cite{Hugle:2018qbw}. As discussed in these earlier works, one can have successful leptogenesis with hierarchical RHN, at a scale as low as 10 TeV, if $N_1$ is assumed to be in thermal bath throughout. Considering freeze-in of $N_1$ from the SM bath (which is more natural for TeV scale $N_1$ due to its tiny Yukawa couplings) pushes the scale of leptogenesis to a slightly higher scale \cite{JyotiDas:2021shi}. Note that, in $N_1$ leptogenesis scenario of minimal scotogenic model, we can not tune $\lambda''_{H \eta}$ arbitrarily to enhance the Yukawa, as this quartic coupling has a lower bound from the requirement of avoiding large inelastic scattering of scalar doublet DM off nucleon.

	Since the scalar singlet can be present in the bath, it can also ensure RHN to be produced in the bath at early epoch irrespective of leptonic Yukawa couplings. Considering the interference between tree level and vertex diagrams, the corresponding expression for the new contribution to CP asymmetry (neglecting singlet scalar mass) can be found as
	\begin{align}
		\epsilon^v_{3 \alpha} &= \frac{1}{32\pi} \frac{\Im(y_{32}\mu_{S\eta} Y^\dagger_{\alpha 3}Y_{\alpha 2})}{Y^\dagger_{\alpha 3}Y_{\alpha 3} M_3} \frac{(r_{N_3}-r_{N_2})}{(r^2_{N_3}-1)^2} \bigg [4r_{N_3}(1  \nonumber \\
		&-r_{N_3}r_{N_2})\log(r_{N_3})+ 2(r^2_{N_3}-1)(r_{N_3} - r_{N_2})\bigg ]
		\label{epsilonflav}
	\end{align}
	where, $r_{N_i} = M_{N_i}/M_\eta$ and we have assumed only $N_2$ be in the loop as $N_1$ coupling with leptons are suppressed. Since we are not going into the resonant regime, we consider only the vertex contribution for simplicity.
	
	\begin{figure}
		\centering
			\includegraphics[scale=0.5]{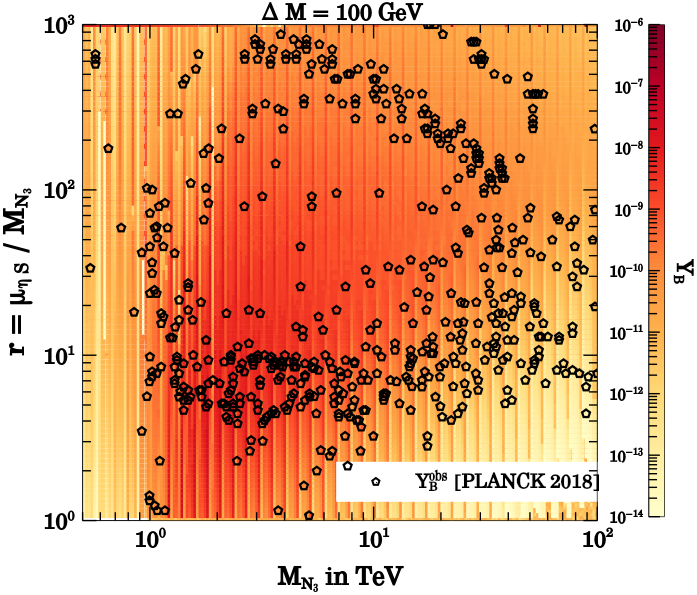}
		\caption{Baryon asymmetry is shown in the plane of $M_{N_{3}}$ and $\mu_{\eta S}/M_{N_{3}}$. Other relevant parameters are fixed at 
			$M_S=1$ GeV, $y_{32}=0.5$, $\lambda''_{H \eta}=0.1$, $m_1=10^{-18}$ eV.}
		\label{leptoyield}
	\end{figure}
	
	We solve the coupled Boltzmann equations by considering the washout as well as annihilation terms mediated by light singlet scalar. For details, please see Appendix~\ref{lepto_appendix}. The $\Delta L=2$ washouts arise due to scatterings $\ell \eta \leftrightarrow \bar{\ell} \eta^*$, $\ell\ell\leftrightarrow \eta^* \eta^*$ mediated by the RHNs and can dominate, specially due to lighter RHNs as mediators. One can also have dominant $\Delta L=1$ scattering processes involving singlet scalar couplings as $\ell N_j \leftrightarrow \eta S $. Since, we are enhancing the trilinear coupling of singlet with $\eta$ to enhance CP asymmetry, we need to tune the corresponding washout involving the same coupling appropriately in order to generate a net asymmetry. This is made possible by choosing a mass spectrum where $N_{2}, \eta$ and $N_3$ masses are in the same ballpark. This introduces Boltzmann suppression to washout processes involving $N_2, \eta$ in external legs. Since DM $N_1$ is light and hence can lead to strong washout processes, we keep its Yukawa coupling with $N_{2,3}$ as well as SM leptons suppressed, resulting in a vanishingly small lightest active neutrino mass. In addition to the washout processes, large annihilation rate of $N_3$ into light singlet scalar can keep it in equilibrium for longer epoch, reducing the asymmetry. However, since $N_3$ coupling with $S$ is not crucial for rest of the phenomenology, we can tune it in such a way to keep $N_3 N_3 \rightarrow S S$ process under control.

	
	We find that the $\Delta L=2$ washout processes remain sub-dominant due to small Yukawa couplings while the $\Delta L=1$ washout processes, specially the one involving trilinear coupling of $S$ with $\eta$ can be very strong. For the compressed mass spectrum chosen as benchmark, these washouts also become Boltzmann suppressed at low temperatures leading to the desired saturation of lepton asymmetry prior to the onset of sphaleron decoupling. If we choose much heavier $N_3$ mass, we can go away from this strong washout regime controlled by $\Delta L=1$ processes. Once the lepton asymmetry gets saturated, its value just prior to the sphaleron temperature ($T_{\rm Sphaleron} \simeq 131$ GeV) can be converted into baryon asymmetry by appropriate conversion factor to generate the observed baryon asymmetry \cite{Aghanim:2018eyx}. In Fig.~\ref{leptoyield}, we have shown the final baryon asymmetry obtained in the plane of $M_{N_{3}}$ and $r=\mu_{\eta S}/M_{N_{3}}$ as $\mu_{\eta S}$ is the key parameter responsible for enhancing the CP asymmetry. Here the black pentagon shaped points represent the parameter space that give rise to correct baryon asymmetry in the observed limit.  

	\section{Conclusion}\label{sec7}
	We have proposed a novel and minimal framework where a light scalar field can give rise to dark matter self-interactions in order to alleviate the tensions of cold dark matter paradigm with small scale structure while simultaneously enhancing the CP asymmetry in RHN decay into leptons providing a feasible leptogenesis scenario even at sub-TeV scale without any fine-tuned mass splitting between RHNs. For demonstration purpose we choose a radiative seesaw scenario although the mechanism is applicable to canonical seesaw scenarios but at the cost of ad-hoc fine-tunings of parameters involved in DM stability. In the current setup, the three RHNs and a scalar doublet, odd under an in-built $Z_2$ symmetry give rise to radiative neutrino mass while the $Z_2$ even singlet scalar give rise to the required velocity dependent self-interactions of DM, assumed to be the lightest RHN $N_1$. While DM abundance remains typically sub-dominant due to strong annihilations into light mediators, late decay of heavier right handed neutrino $N_2$ can fill up this deficit. The same singlet scalar as well as $N_2$ also appear in one loop decay of the heaviest RHN $N_3$ into SM leptons providing a large CP asymmetry due to the interference with the tree level decay. While trilinear coupling of singlet scalar with $Z_2$-odd scalar doublet enhances the CP asymmetry even for sub-TeV scale $N_3$, its trilinear coupling with the SM Higgs can create the scalar portal for DM-nucleon scattering keeping the scenario predictive at direct detection experiments. The same mixing of such light scalar with the SM Higgs can also be probed at colliders via Higgs invisible decay, displaced vertex as well as meson decays \cite{Gershtein:2020mwi}. Also, the $Z_2$-odd sector particles can lead to missing transverse energy signatures \cite{Miao:2010rg, Gustafsson:2012aj, Belyaev:2016lok, Belyaev:2018ext} as well as displaced vertex signatures \cite{Borah:2018smz}. Additionally, due to the requirement of keeping the DM induced washouts of lepton asymmetry sub-dominant, the corresponding Yukawa coupling of $N_1$ with leptons remain suppressed leading to a vanishingly small lightest active neutrino mass. On the other hand, the heavier RHNs can have large Yukawa couplings with SM leptons and hence can give rise to observable charged lepton flavour violation. For $N_{2,3}, \eta$ masses around the TeV ballpark as chosen here, the model can saturate MEG bound on $\mu \to e \gamma$ \cite{TheMEG:2016wtm} for $\lambda''_{H \eta} < 10^{-6}$.

	\acknowledgements
	NS acknowledges the support from Department of Atomic Energy (DAE)- Board of Research in Nuclear Sciences (BRNS), Government of India (Ref. Number: 58/14/15/2021- BRNS/37220).
	
	
	\appendix
	\section{Boltzmann equation for DM production}\label{dm_BE}
	
	\begin{figure*}[htb!]
		\centering
		\begin{tabular}{cc}
			\includegraphics[height=7cm,width=7.5cm]{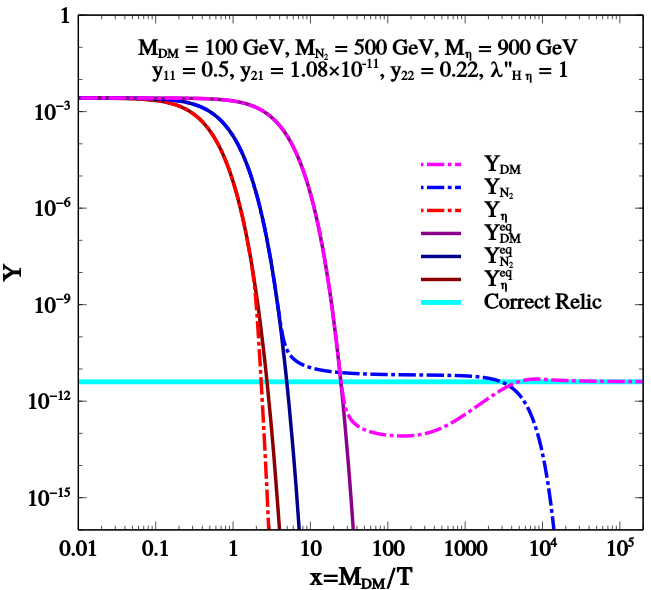}
			\includegraphics[height=7cm,width=7.5cm]{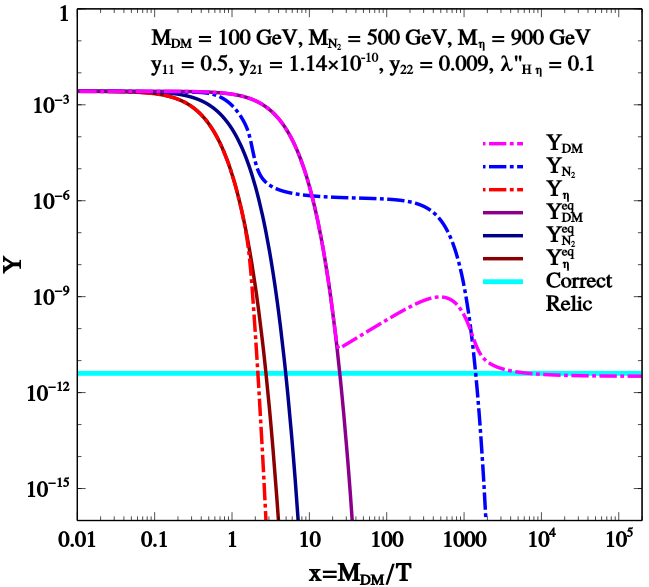}
		\end{tabular}
		\caption{ Evolution of different $Z_2$-odd particles relevant for DM relic generation for a fixed benchmark set of parameters. The values of the parameters used are; (i)[Left Panel]: $M_{DM} = 100$ GeV, $M_{N_{2}} = 500$ GeV, $M_{\eta} = 900$ GeV, $y_{11} = 0.5$, $y_{21} = 1.08\times10^{-11}$, $y_{22} = 0.22,~ \lambda''_{H \eta} = 1$ and (ii)[Right Panel]: $M_{DM} = 100$ GeV, $M_{N_{2}} = 500$ GeV, $M_{\eta} = 900$ GeV, $y_{11} = 0.5$, $y_{21} = 1.14\times10^{-10}$, $y_{22} = 0.009,~ \lambda''_{H \eta} = 0.1$.}
		\label{dm_relic}
	\end{figure*}
	
	The relevant Boltzmann equations for DM as well as $N_2, \eta$ in terms of their comoving number densities are given by 
		\begin{widetext}
			\begin{equation}
				\label{eq:BE1}
				\begin{aligned}
					\frac{dY_{N_{{1}}}}{dx}&=-\frac{s(M_{\rm DM})}{x^2  H(M_{\rm DM})}\big( \langle \sigma(N_{{1}} N_{{1}} \to {\rm All}) v \rangle \big) (Y^2_{N_1} -\big(Y^{\rm eq}_{N_1}\big)^2) +\frac{x}{H(M_{\rm DM})}\big( \langle \Gamma_{N_2 \rightarrow S N_1}\rangle Y_{N_2} + \langle \Gamma_{N_2 \rightarrow N_1ll}\rangle Y_{N_2} + \langle \Gamma_{\eta \rightarrow N_1l}\rangle Y_{\eta} \big),\\
					\frac{dY_{N_2}}{dx}&=-\frac{s(M_{\rm DM})}{x^2  H(M_{\rm DM})}\big( \langle \sigma(N_2 N_2 \to {\rm All}) v \rangle \big) (Y^2_{N_2} -\big(Y^{\rm eq}_{N_2}\big)^2)+\frac{x}{H(M_{\rm DM})}\big( \langle \Gamma_{\eta \rightarrow N_2l}\rangle Y_{\eta} - \langle \Gamma_{N_2 \rightarrow N_1ll}\rangle Y_{N_2}-\langle \Gamma_{N_2 \rightarrow N_1S}\rangle Y_{N_2}\big),\\
					\frac{dY_{\eta}}{dx}&=-\frac{s(M_{\rm DM})}{x^2  H(M_{\rm DM})}\big( \langle \sigma(\eta \eta \to {\rm All}) v \rangle \big) (Y^2_{\eta} -\big(Y^{\rm eq}_{\eta}\big)^2)-\frac{x}{H(M_{\rm DM})}\big( \langle \Gamma_{\eta \rightarrow N_{i}l}\rangle Y_{\eta} \big).
				\end{aligned}
			\end{equation}
		\end{widetext}
		In the above equations, $x=\frac{M_{\rm DM}}{T}$, $s(M_{\rm DM})= \frac{2\pi^2}{45}g_{*s}M^3_{\rm DM}$ , $H(M_{\rm DM})=1.67 g^{1/2}_*\frac{M^2_{\rm DM}}{M_{\rm Pl}}$ and $<\sigma(A A \to {\rm All}) v>$ represents the thermally averaged cross-section for annihilation of $A$ ($A=N_1,N_2,\eta$) to all possible final state particles.

Solving these coupled Boltzmann equations numerically, we show the evolution of relevant species for two different benchmark points in the left and right panel of Fig. \ref{dm_relic}. In both the figures, the evolution of comoving number density of $DM$, $N_2$ and $\eta$ are shown by the magenta, blue and red dot-dashed lines respectively. The solid lines represents their equilibrium number density. For both the benchmark points we have fixed the masses as: $M_{DM}=100$ GeV, $M_{N_2}=500$ GeV and $M_{\eta}=900$ Gev and the Yukawa coupling for self-interaction of DM $y_{11}=0.5$, whereas the couplings $y_{22}$ and $y_{21}$ are different. 
	For the first benchmark point, $y_{22}$ is large ($y_{22}=0.22$), so the DM freeze-out abundance is small and thus $y_{21}$ is tuned to $y_{21}=1.08\times10^{-11}$ such that the DM gets a non-thermal contribution from $N_2$ decay, after its thermal freeze-out, to fill the deficit which is depicted in the left panel of Fig.~\ref{dm_relic}. For the second benchmark point, shown in the right panel of Fig.~\ref{dm_relic}, $y_{22}$ is smaller ($y_{22}=0.009$), and hence it decouples from the thermal bath with a large abundance. Thus DM  should get the non-thermal contribution to its abundance from an earlier epoch due to a larger Yukawa coupling with $N_2$ ($y_{21}=1.14\times10^{-10}$) such that the final DM abundance settles down to the required value due to subsequent annihilations into light mediators. 
	
	\section{Dark Matter Self-interactions}\label{self_int}
The non-relativistic DM self-scatterings are described in terms of the transfer cross-section $\sigma_T$ as~\cite{Feng:2009hw,Tulin:2013teo,Tulin:2017ara}
		\begin{equation}
			\sigma_T = \int d\Omega (1-\cos\theta) \frac{d\sigma}{d\Omega}
		\end{equation}
		In the Born Limit ($y^2_{11} M_{\rm DM}/(4\pi M_S) << 1$),
		\begin{equation}
			\sigma^{\rm Born}_T=\frac{y^4_{11}}{2\pi M^2_{\rm DM} v^4}\Bigg(\ln(1+\frac{ M^2_{\rm DM} v^2}{M^2_S})-\frac{M^2_{\rm DM}v^2}{M^2_S+ M^2_{\rm DM}v^2}\Bigg)
		\end{equation} 
		In the classical regime ($y^2_{11} M_{\rm DM}/4\pi M_S\geq 1, M_{\rm DM} v/M_{S} \geq 1$)~\cite{Tulin:2013teo}:
		\begin{equation}
			\sigma^{\rm Classical}_T =\left\{
			\begin{array}{l}		
				\frac{4\pi}{M^2_S}\beta^2 \ln(1+\beta^{-1}) ~~~~~~~~~~~\beta \leqslant 10^{-1}\\
				\frac{8\pi}{M^2_S}\beta^2/(1+1.5\beta^{1.65}) ~~~~~~~~10^{-1}\leq \beta \leqslant 10^{3}\\
				\frac{\pi}{M^2_S}( \ln\beta + 1 -\frac{1}{2}\ln^{-1}\beta) ~~~~~\beta \geq 10^{3}\\
			\end{array}
			\right.
		\end{equation}  
		where $\beta = y^2_{11} M_{\rm DM}/(2\pi M_S) v^2$.
	
In the resonant regime for ($y^2_{11} M_{\rm DM}/(4\pi M_S) \geq 1, M_{\rm DM} v/M_{S} \leq 1$), there is no analytical formula for $\sigma_T$ and 
		the non-perturbative results obtained by approximating the Yukawa potential to a Hulthen potential $\Big( V(r) = \pm \frac{y^2_{11}}{4\pi}\frac{ \delta e^{-\delta r}}{1-e^{-\delta r}}\Big)$ is used, which is given by~\cite{Tulin:2013teo}:
		\begin{equation}
			\sigma^{\rm Hulthen}_T = \frac{16 \pi \sin^2\delta}{M^2_{\rm DM} v^2}
		\end{equation}
		where phase shift $\delta$ is given in terms of the $\Gamma$ functions by :
		\begin{eqnarray}
			\delta &={\rm arg} \Bigg(i\Gamma \Big(\frac{i M_{\rm DM} v}{k~ M_S}\Big)\bigg/{\Gamma (\lambda_+)\Gamma (\lambda_-)}\Bigg)\nonumber\\
			\lambda_{\pm} &=
			\begin{array}{l}
				1+ \frac{i M_{\rm DM} v}{2 ~k ~M_S}  \pm \sqrt{\frac{y^2_{11}M_{\rm DM}}{4 \pi k M_S} - \frac{ M^2_{\rm DM} v^2}{4 k^2 M^2_S}}\\
			\end{array}
		\end{eqnarray}   
		and $k \approx 1.6$ is a dimensionless number. 
	
	\section{Boltzmann equations for Leptogenesis}\label{lepto_appendix}
\begin{figure}[htb!]
		\centering
		\includegraphics[scale=0.28]{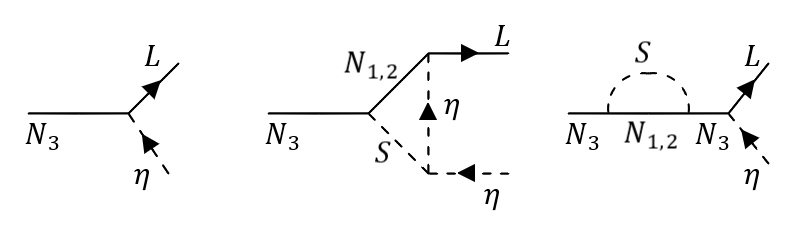}
		\caption{Leptogenesis from $N_3$ decay with dominant contribution to CP asymmetry from singlet scalar couplings.}
		\label{lepto}
	\end{figure}

	The Feynman diagrams corresponding to the dominant source of CP asymmetry generation for TeV scale leptogenesis are shown in Fig. \ref{lepto}.
	
The Boltzmann equations for comoving number densities of $N_3$ and $B-L$ are given by:
		\begin{eqnarray}
			\frac{dY_{N_3}}{dz}&  \ = \ & -(D_3+D_{32}+D_{31}) (Y_{N_3}-Y_{N_3}^{\rm eq}) \nonumber \\
			&+& \sum_{j=1,2} D_{3j} (Y_{N_j}-Y_{N_j}^{\rm eq})-\sum_{j=1,2,3} \Delta_{N_3 N_j} S_{N_3 N_j}\nonumber\\
			\label{eq:bol1}
		\end{eqnarray}
		\begin{eqnarray}
			\frac{dY_{B-L}}{dz}& \ = \ &-\epsilon_3 D_3 (Y_{N_3}-Y_{N_3}^{\rm eq})-W^{\rm Total}Y_{B-L} \, 
			\label{eq:bol2}
		\end{eqnarray}
		where $Y_{N_3}^{\rm eq}=\frac{z^2}{2}K_2(z)$ is the equilibrium comoving number density of $N_3$ (with $K_i(z)$ being the modified Bessel function of $i$-th kind). The quantity on the right hand side of the above equations
		\begin{align}
			D_3 \ \equiv \ \frac{\Gamma_{3}}{Hz} \ = \ K_{N_3} z \frac{K_1(z)}{K_2(z)}, \nonumber \\
			D_{3j} \ \equiv \ \frac{\Gamma_{3j}}{Hz} \ = \ K_{N_3, j} z \frac{K_1(z)}{K_2(z)}
			\label{D1}
		\end{align}
		measures the partial decay rates of $N_3$ with respect to the Hubble expansion rate as
		$$ K_{N_3}=\frac{\Gamma (N_3 \rightarrow L \eta)}{H}, \, K_{N_3, j}=\frac{\Gamma (N_3 \rightarrow N_j S)}{H}. $$
		The third term on the right hand side of equation \eqref{eq:bol1} are defined as
		$$ \Delta_{N_i N_j} = (Y_{N_i} Y_{N_j} - Y^{\rm eq}_{N_i} Y^{\rm eq}_{N_j})/(Y^{\rm eq}_{N_i} Y^{\rm eq}_{N_j})$$ 
		whereas $S_{N_i N_j}$ is the thermally averaged scattering cross-section. Similarly, $W^{\rm Total} \equiv \frac{\Gamma_{W}}{Hz}$ measures the total washout rate.The washout term is the sum of two contributions, coming from $\Delta L=1$ and $\Delta L=2$ washouts. One of the major $\Delta L=1$ process is the inverse decay of leptons and $\eta$. The other contribution to washout $ W^{\rm Total}$ originates from scatterings which violate lepton number by $\Delta L=1, 2$.

		\begin{table}[h!]
			\centering
			\begin{tabular}{|c|c|c|c|}
				\hline
				BP &  $M_{N_{3}}$ (TeV) & $\Delta M$ (GeV) & $\mu_{S \eta}/M_{N_{3}}$ \\
				\hline
				BP1 & 2 & 100 & 7 \\
				BP2 & 4 & 200 & 5 \\
				BP3 & 8 & 500 & 5 \\
				\hline
			\end{tabular}
			\caption{Benchmark values used in Fig. \ref{leptoyield}.}
			\label{tab:1}
		\end{table}
		\begin{figure}
			\centering
				\includegraphics[scale=0.5]{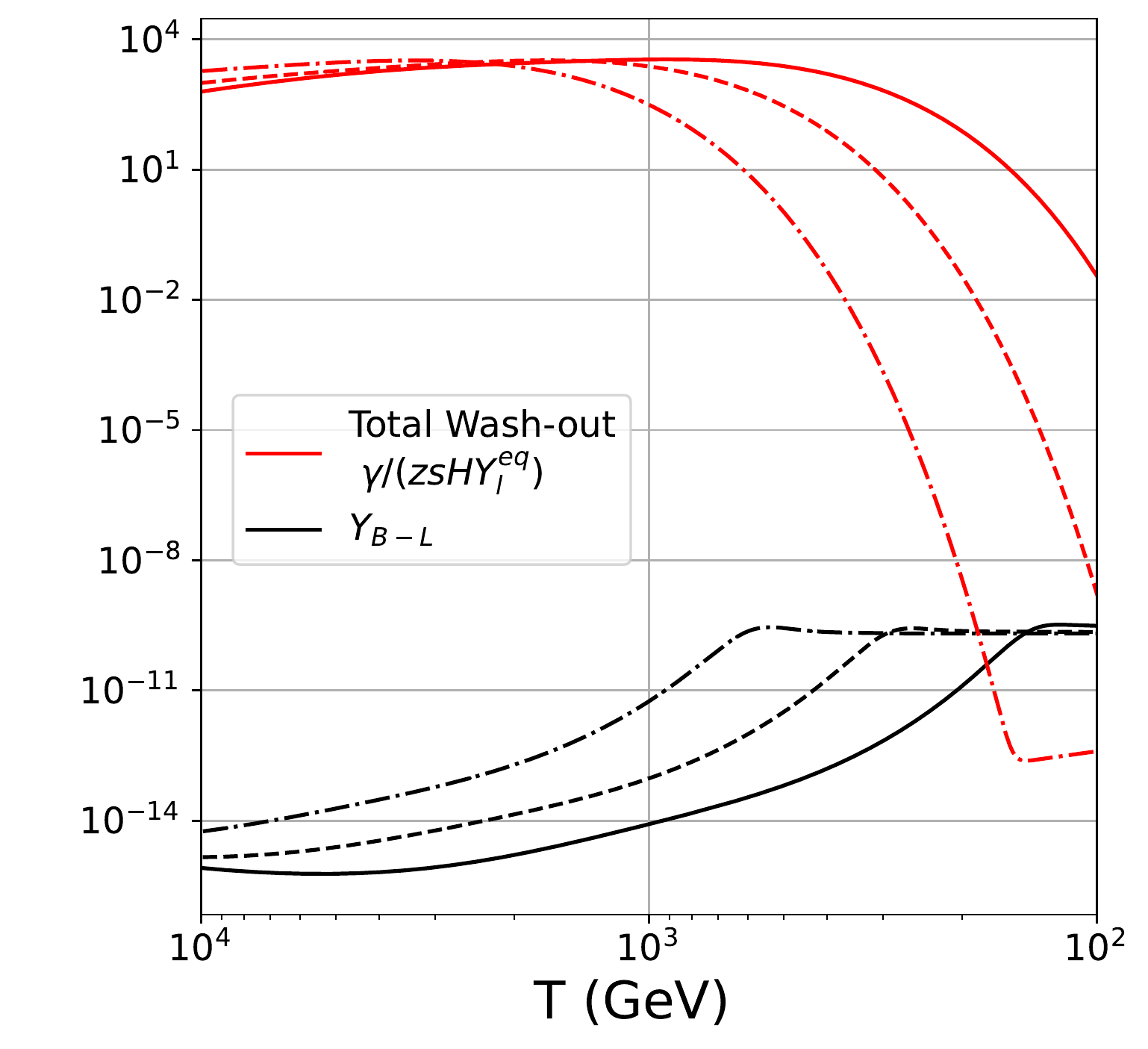}
			\caption{Evolution of lepton asymmetry and washout rates for benchmark values BP1 (solid), BP2 (dashed), BP3 (dot-dashed) shown in Table \ref{tab:1}.}
			\label{leptoyield}
		\end{figure}
		The Dirac Yukawa couplings of neutrinos are written in Casas-Ibarra parametrisation \cite{Toma:2013zsa} as
		\begin{equation}\label{Yukawa}
			Y = {\sqrt{\Lambda}}^{-1} R \sqrt{\hat{m_\nu}} U^\dagger_{\rm PMNS} \end{equation}
		where $R$, in general, is an arbitrary complex orthogonal matrix satisfying $RR^{T}=\mathrm{I}$ that can be parametrised in terms of three complex angles $z_{ij}$.  Here, $\hat{m_\nu} =  \textrm{Diag}(m_1,m_2,m_3)$ is the diagonal light neutrino mass matrix and the diagonal matrix $\Lambda$ is defined as $\Lambda$ = Diag ($\Lambda_1$,$\Lambda_2$,$\Lambda_3$), with
		$\Lambda_i$'s being related to the loop factor. In Eq. (\ref{Yukawa}), 
		$U_{\rm PMNS}$ represents the usual Pontecorvo-Maki-Nakagawa-Sakata (PMNS) mixing matrix of neutrinos. 
	
	The evolution of comoving densities of lepton number and different washout processes are shown in Fig. \ref{leptoyield} for benchmark parameters listed in caption as well as in Table \ref{tab:1}. We have ignored $N_3$ self-coupling with $S$ namely $y_{33}$ for simplicity and the orthogonal matrix $R$ is parametrised as a rotation in 2-3 plane with complex mixing angle $z_{23}=\pi/4-i\pi/2$. We also keep the mass splitting $\Delta M$ between $N_3$ and $N_2, \eta$ to be same, for simplicity. The interplay of washout processes and asymmetry evolution is clearly visible in the plot. In order to show the washout rates, we plot the dimensionless quantity $\gamma (\Delta L=1, 2)/(sHz Y^{\rm eq}_l)$ where $\gamma = \frac{T}{64\pi^2} \int^{\infty}_{s_0} ds \hat{\sigma}(s) \sqrt{s} K_1(\sqrt{s}/T)$ with $\hat{\sigma}$ being the reduced cross section \cite{Hambye:2001eu}.


	\twocolumngrid
	%

\end{document}